\documentclass[]{emulateapj}
\shorttitle{Photometric redshift of the GRB 981226 host galaxy}
\shortauthors{Christensen et al.}

\begin{document}

\title{Photometric redshift of the GRB 981226 host galaxy 
   \thanks{Based on
    observations made with the European Southern Observatory telescopes
    obtained from the ESO/ST-ECF Science Archive Facility (Programme IDs:
    67.B-0611(A), 66.B-0539(A), 265.D-5742(A), 165.H-0464(C), 265.D-5742(C),
    67.D-0273(A), and HST proposal ID 8640).}  }

\author{L. Christensen\altaffilmark{1}}
\affil{Astrophysikalisches Institut Potsdam, An der Sternwarte
  16,  14482 Potsdam, Germany}
\email{lchristensen@aip.de}

\author{J. Hjorth\altaffilmark{2}}
 \affil{Dark Cosmology Centre, Niels Bohr
Institute, University of Copenhagen, Juliane Maries
    Vej 30, DK--2100 Copenhagen, Denmark}
\email{jens@astro.ku.dk}

\and
\author{J. Gorosabel\altaffilmark{3}}
\affil{Instituto de Astrof\'{\i}sica de Andaluc\'{\i}a, IAA-CSIC, P.O. Box
  03004, E-18080 Granada,  Spain}
\email{jgu@iaa.es}

\begin{abstract}
  No optical afterglow was found for the dark burst \object{GRB 981226} and
  hence no absorption redshift has been obtained.  We here use ground-based
  and space imaging observations to analyse the spectral energy distribution
  (SED) of the host galaxy. By comparison with synthetic template spectra we
  determine the photometric redshift of the GRB 981226 host to be
  $z_{\mathrm{phot}}$ = 1.11$\pm$0.06 (68\% confidence level). While the
  age-metallicity degeneracy for the host SED complicates the determination of
  accurate ages, metallicity, and extinction, the photometric redshift is
  robust.  The inferred $z_{\mathrm{phot}}$ value is also robust compared to a
  Bayesian redshift estimator which gives $z_{\mathrm{phot}}=0.94\pm0.13$.
  The characteristics for this host are similar to other GRB hosts previously
  examined.  Available low resolution spectra show no emission lines at the
  expected wavelengths. The photometric redshift estimate indicates an
  isotropic energy release consistent with the Amati relation for this GRB
  which had a spectrum characteristic of an X-ray flash.
\end{abstract}


\section{Introduction}

The Swift satellite \citep{gehrels04} promises the build-up of a significant
sample of gamma-ray bursts (GRBs) with well-understood selection criteria
useful for cosmological studies of high-redshift galaxies \citep{jakobsson05b}
and the Hubble diagram \citep{ghirlanda04}. The best way of securing GRB
redshifts is from absorption line studies of the afterglow.  However, for the
first 56 Swift GRBs only 11 redshifts have been obtained.  Alternatively,
redshifts can be obtained from host galaxy spectroscopy although this requires
that the host is sufficiently bright and has detectable emission lines.

In this Letter we explore a third approach suitable for fainter host galaxies
in case a spectroscopic absorption or emission line redshift has not been
obtained.  For a sample of 10 GRB host galaxies with spectroscopic redshifts
in the range $0.4<z<2$ and photometric measurements in more than 4 bands,
\citet{lise04b} found that photometric redshifts were consistent with the
spectroscopic redshifts in all cases, within $\Delta z = 0.21$. An advantage
of the method is that it is independent of whether the host galaxy has
emission lines.

As the next step in validating the photometric redshift approach we here
predict a redshift from a host galaxy with multiband photometry which is
sufficiently bright that a spectroscopic redshift can be measured, and thereby
test the photometric redshift. Our target is \object{GRB~981226} which is
currently one of the brightest hosts without a spectroscopic redshift.

GRB 981226 was detected by the BeppoSAX satellite on 1998 December 26.41 UT.
It is consistent with being an X-ray flash (XRF) with the fluences satisfying
$S_X > S_{\gamma}$ \citep{frontera00}.  Despite intense optical and
near-infrared (IR) follow-up observations initiated 6.5 and 8.4 hours after
the burst, respectively \citep{gcn186,gcn173} no optical counterpart was found
\citep{gcn172,gcn177,gcn181,gcn182,gcn185}.  The deep $R$-band observations
carried out by \citet{gcn190} showed $R > 23$ mag at 9.9 hours after the
GRB. This makes GRB 981226 a typical GRB without any detected afterglow
\citep{taylor98} and close to being a dark burst \citep{jakobsson04,rol05}.

Radio observations revealed a variable source at the position
R.A.(J2000)=23$^h$29$^m$37\fs21, Dec(J2000)=$-23^\circ$55\arcmin53\farcs8
peaking $\sim$10 days after the gamma-ray event \citep{frail99}.
Identification of the host galaxy was suggested based on the small angular
separation between the radio afterglow and an extended object.  High spatial
resolution images from the HST/STIS show that the galaxy colours change
notably over its surface, with the northern part being significantly bluer
\citep{holland_gcn}.  The position of the radio afterglow with respect to the
galaxy is 0\farcs749$\pm$0\farcs328 \citep{bloom00}, which encompasses
the blue northern part of the host.

This Letter presents an analysis of all imaging data on the host available in
public archives. Ground-based data in the $BV\!RIJ\!sK\!s$ bands and images
from the HST makes a multi-wavelength study appropriate to determine the host
and hence the GRB photometric redshift.  Some of the data have been presented
previously in the literature in different contexts. An ISAAC $K\!s$ band image
was presented in \citet{lefloch03}, the HST/STIS data in \citet{holland_gcn},
and a deep $R$ band image in \citet{frail99}; the latter image is not analysed
here.

\section{Data analysis}

For consistency we re-analysed the photometry in all filters.  The host galaxy
was observed in the optical ($BV\!RI$) with VLT/FORS1 and in the near-IR
($J\!s$ and $K\!s$) with ISAAC. The data were retrieved from ESO's public
archive\footnote{http://archive.eso.org}.  To take advantage of all available
data on the host we also included images from the Hubble Space Telescope using
STIS.  The host was observed in July 2000, using a clear aperture (50CCD, or
$C\!L$ filter) and a long pass imaging filter, F28$\times$50LP ($LP$
filter)\footnote{Images were obtained from
  http://www.ifa.au.dk/$^{\sim}$hst/grb\_hosts/intro.html \citep{holland01q}.
  The data are part of the Cycle 9 programme GO-8640 ``A Public Survey of Host
  Galaxies of Gamma-ray Bursts''.}. Table~\ref{tab:log} lists the dates of
observations, number of integrations and exposure times.

\begin{deluxetable}{lll}
%\tablewidth{0pt}
\tablecaption{Log of the observations \label{tab:log}}
\tablehead{
\colhead{Instrument+filter} & \colhead{date} & \colhead{exp time (s)}
}
\startdata
  FORS1 $R$ & 2000-10-05 & 18$\times$540\\
  FORS1 $B$ & 2001-05-15 & 3$\times$300\\
  FORS1 $V$ & 2001-05-16 & 3$\times$300\\
  FORS1 $R$ & 2001-06-19 & 6$\times$540\\
  FORS1 $I$ & 2001-05-16 & 3$\times$300\\
  FORS1 $I$ & 2001-08-13 & 400\\
  FORS1 $I$ & 2001-08-19 & 400\\
  FORS1 $I$ & 2001-09-22 & 4$\times$600\\
  ISAAC $J\!s$& 2001-09-22 & 10$\times$180\\
  ISAAC $K\!s$& 2000-11-12 & 30$\times$120\\
  ISAAC $K\!s$& 2001-06-07 & 30$\times$120\\
  STIS $C\!L$& 2000-07-03  & 8265$^a$\\
  STIS $LP$& 2000-07-06    & 7909$^a$\\ 
\enddata
\tablenotetext{a}{Total integration time.}
\end{deluxetable}

Data reduction was done using IRAF.  All optical data were reduced using
standard methods, i.e. bias subtraction and dividing by a combined average
flat-field frame obtained from twilight sky exposures.  For the near-IR data
reduction, sky subtraction was done by creating a sky image from the
bracketing 8 frames from each individual night. After sky subtraction each
image was divided by a normalised flat-field frame.  Near-IR flat-field images
were created by subtracting faintly illuminated flat-field frames from bright
ones.  Eight and ten of such images were combined and used for flat-fielding
the $J\!s$ and $K\!s$ images, respectively.  Relative shifts were found using
cross-correlation procedures before combining the individual exposures.

Additional archive calibration files consisting of bias, sky flat-field, and
standard star images from the different nights were retrieved. Photometric
zero-points from ISAAC and FORS1 were obtained by comparing the instrumental
magnitude of the standard star with tabulated values. All the zero-points
calculated confirmed the values for the given dates that were found in the
corresponding instrument web-pages. Extinction and colour terms are reported
for each month on the FORS1 web page and were assumed to be appropriate here.
For the STIS data, zero-points were obtained from the STIS user manual.

Aperture photometry was used to derive host photometry, and all magnitudes of
the host listed in Table~\ref{tab:mags} were derived using a 2\arcsec\ radial
aperture.  Corrections to larger apertures were found to be negligible.
No optical afterglow was found for this burst and the observations were
carried out 1.5--2.5 years after the burst. Therefore, no contamination of any
significant level is expected for the photometry of the host.  The magnitudes
derived here are consistent with $R=24.85\pm0.06$ mag in \citet{frail99},
$K\!s=21.1\pm0.2$ mag in \citet{lefloch03}, and $C\!L=25.04\pm0.07$ mag in
\citet{holland_gcn} (using the same 1\farcs1 radial aperture we find
25.00$\pm$0.05 mag).

\begin{deluxetable}{lll}
\tablewidth{0pt}
\tablecaption{Host galaxy magnitudes \label{tab:mags}}
\tablehead{
\colhead{Filter} & \colhead{$m_{\mathrm{Vega}}$(mag)} & \colhead{$m_{\mathrm{AB}}$(mag)}
}
\startdata
  $B$ & 25.74$\pm$0.37 & 25.68$\pm$0.37\\ 
  $V$ & 25.59$\pm$0.15 & 25.63$\pm$0.15\\
  $R$ & 24.81$\pm$0.06 & 25.04$\pm$0.06\\
  $I$ & 23.65$\pm$0.15 & 24.09$\pm$0.15\\
  $J\!s$& 21.90$\pm$0.20& 22.84$\pm$0.20\\
  $K\!s$& 21.10$\pm$0.27& 22.97$\pm$0.27\\
  $C\!L$&               & 24.79$\pm$0.10\\
  $LP$&                 & 24.27$\pm$0.10\\
\enddata
\tablecomments{Magnitudes of the host in the Vega and AB systems in
  various filters obtained with an aperture radius of 2\arcsec. No
  correction for Galactic extinction is applied.}
\end{deluxetable}

%__________________________________________________________________

\section{Photometric redshift}

We used the public photometric redshift code HyperZ \citep{bolzo00} along
similar lines as in a series of papers on GRB host galaxies
\citep{goro03b,goro03,lise04,goro05}. As shown in these papers, photometric
redshifts of GRB hosts can be determined to within $\Delta z=0.21$ when
multiband observations are available, and when spectral features such as the
Balmer jump are bracketed by the observations.

Magnitude offsets between Vega and the AB system were calculated and added to
the observed Vega magnitudes listed in Table~\ref{tab:mags}. The HST
magnitudes are obtained directly in the AB system (see STIS Instrument
Handbook), so the Table~\ref{tab:mags} fields corresponding to the HST Vega
magnitudes are empty. To derive fluxes for the various pass bands the
magnitudes were corrected for a Galactic extinction of $E(B-V)$ = 0.022 mag
as derived from the dust maps of \citet{schlegel98}. Flux densities in $\mu$Jy
were calculated by \(f_{\nu} = 10^{-0.4(m_{\mathrm{AB}}-23.9)}\) for an AB
magnitude, $m_{\mathrm{AB}}$, in each passband.

Fluxes were compared to galaxy template spectra created from the spectral
atlas of \citet{bru93}. We used different galaxy templates obtained from a
Salpeter initial mass function (IMF) \citep{salpeter55} (Sp55) and a Miller \&
Scalo IMF \citep{miller79} (MiSc79) with different star formation histories
and a characteristic timescale $\tau$, where $\tau=0$ corresponds to a
starburst template, and $\tau\rightarrow \infty$ to an irregular galaxy
template. Values of $\tau$ in between these extremes correspond to various
types of spiral galaxy and elliptical templates.  These templates all assume
solar metallicity. We also experimented using different extinction curves: a
starburst extinction curve \citep{calz00}, a Small Magellanic Cloud (SMC)
extinction curve \citep{prevot84}, a Large Magellanic Cloud (LMC) extinction
curve \citep{fitz86}, and the Milky Way extinction curve \citep{seaton79}.

In principle the STIS wide filters might be included to determine the
photometric redshift. The $C\!L$ filter extends from 4000 to 9000 {\AA} and
the $LP$ from 5500 to 9000 {\AA}, thus providing poor spectral information.
Furthermore given the widths of the filters, the effective wavelength is
sensitive to the assumed spectral template. Thus, as a first step, we decided
to not include them in our analysis, and then discuss the impact that their
inclusion has on the $z_{\mathrm{phot}}$ determination.

\begin{figure}
\plotone{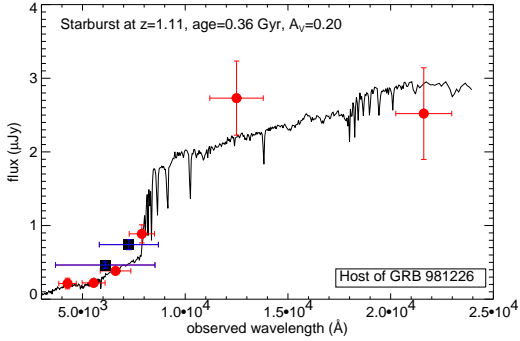} 
\caption{Best fit synthetic spectrum to the SED of the GRB 981226 host
    galaxy using the ground based $BVRIJ\!sK\!s$ images. The template was
    created from a Sp55 IMF and a Calzetti extinction law.  The broad STIS
    filters are indicated by the squares, but are not included in the fit.
    Horizontal error bars indicate the widths of each band.\label{fig:fit}}
\end{figure}

The best fit of the broad band SED was found with a starburst template at
$z=1.11$ with an age of 0.36 Gyr and an intrinsic extinction of $A_V=0.20$ mag
as shown in Fig.~\ref{fig:fit}. Using a Calzetti extinction curve and a Sp55
IMF gave the smallest reduced $\chi^2$/d.o.f. as indicated in
Table~\ref{tab:list}.  Other combinations of template fits are also listed in
Table~\ref{tab:list}.  In all cases a starburst template provided the best
  fit. From these fits we find $z_{\mathrm{phot}}$ =
1.11$\pm$0.06$\pm$0.10$\pm$0.21 (68\%, 90\%, and 99\% confidence levels,
respectively). In this case the Balmer jump is well sampled by the
  observations.

\begin{deluxetable}{lllll}
  \tablewidth{0pt} \tablecaption{Solar metallicity SED fits \label{tab:list}}
  \tablehead{ \colhead{IMF} & \colhead{ext. law} &
    \colhead{$z_{\mathrm{phot}}$ (68\%,99\%)}& \colhead{$A_V$}(mag) &
    \colhead{$\chi^2$/d.o.f.}  }  \startdata
  Sp55  & Cal00&  1.11$^{+0.06,+0.21}_{-0.04,-0.18}$ &  0.20 & 0.399\\
  MiSc79 & Cal00& 1.11$^{+0.06,+0.20}_{-0.06,-0.22}$ &  0.10 & 0.661\\
  Sp55  & SMC   & 1.11$^{+0.07,+0.23}_{-0.03,-0.45}$ &  0.12 & 0.427\\
  MiSc79 & SMC& 1.11$^{+0.06,+0.21}_{-0.05,-0.21}$ &  0.06 &  0.651\\
  Sp55   & LMC& 1.11$^{+0.07,+0.23}_{-0.04,-0.46}$ &  0.16 &  0.426\\
  MiSc79 & LMC& 1.11$^{+0.06,+0.21}_{-0.06,-0.22}$ &  0.08 &  0.664\\
  Sp55   & MW & 1.11$^{+0.07,+0.24}_{-0.04,-0.17}$ &  0.14 &  0.432\\
  MiSc79 & MW & 1.11$^{+0.06,+0.21}_{-0.05,-0.21}$ &  0.06 &  0.662 \\
  \enddata \tablecomments{Template SED fits using a Salpeter or Miller \&
    Scalo IMFs, and different extinction curves. A redshift step of $\Delta z
    =0.05$ was used in our SED fits.  Only ground-based photometric points
    were included in the fits.  The derived properties in terms of best fit
    template type (Starburst), redshift, age (0.36 Gyr), $M_B=-20.25$,
    and extinction are basically independent of the input parameters for the
    assumed IMF and extinction curve.  A relative host luminosity
    $L/L^*\approx0.5$ is derived assuming $M_B^*=-21$ mag.}
\end{deluxetable}

\begin{figure}
\plotone{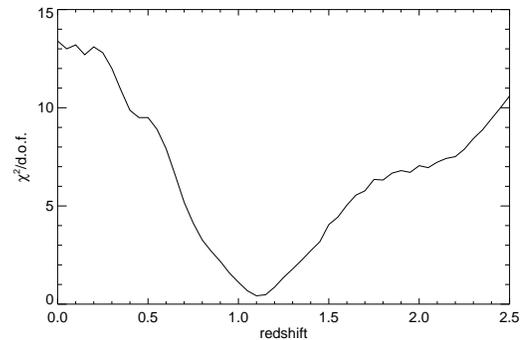}
\caption{Reduced $\chi^2$ as a function of redshift with a minimum at
  $z\approx1.11$ for the fit using the ground based data only. \label{fig:zphot}}
\end{figure}

The photometric redshift determination has a well defined minimum at
$z\approx1.1$ as shown in Fig.~\ref{fig:zphot}.  The best fit values for the
extinction, age, and redshifts are consistent independently of the extinction
curve or template used. Using the observed template spectra from
\citet{kinney96} give consistent results for the photometric redshift and
extinction.  Including also the HST bands in the SED fit, the resulting
parameters ($M_B$, age, $A_V$) do not change.  Specifically
$z_{\mathrm{phot}}$ does not change at all, but the uncertainty for the
photometric redshift decreases by $\sim$50\%.  However, including the HST
photometry increases the $\chi^2$. A fit when including the $LP$ band gives
$\chi^2/\mathrm{d.o.f.}=1.315$, while including the $C\!L$ as well gives
$\chi^2/\mathrm{d.o.f.}=2.850$. The reason for this is the calculation of the
effective wavelength of the two HST/STIS bands.  In Fig.~\ref{fig:zphot} the
HST data points are shown by the squares.

We investigated whether using templates with metallicities different from
solar values would have an impact on the output parameters. It is expected
that the photometric redshift shows no significant difference while the
well-known age-metallicity degeneracy would manifest itself. In
Table~\ref{tab:metal} we list the results of fitting the observed SED with
templates of metallicities of 0.2 and 0.4 solar, respectively. The template
spectra were calculated using a Sp55 IMF with the GALAXEV code
\citep{bruzual03}.  Generally the best fit templates have larger ages, while
the photometric redshifts are in good agreement with those in
Table~\ref{tab:list}. Lower metallicities as well as templates constructed
with a \citet{chabrier03} IMF give consistent results for $z_{\mathrm{phot}}$.
Since all the fits give similar values of $\chi^2$, we can not disentangle
this age-metallicity-extinction degeneracy using broad-band measurements alone
\citep[see also][]{bolzo00}.

\begin{deluxetable}{lllll}
\tablewidth{0pt}
\tablecaption{Sub-solar metallicity SED fits. \label{tab:metal}}
\tablehead{
\colhead{Z/Z$_{\sun}$}          & \colhead{ $z_{\mathrm{phot}}$ (68\%,99\%)}  &
\colhead{Age (Gyr)}          & \colhead{$A_V$ (mag)}    &
\colhead{$\chi^2$/d.o.f.}}
\startdata
0.2 & 1.11$^{+0.09,+0.27}_{-0.06,-0.22}$  & 0.72 & 0.00 & 0.476 \\
0.4 & 1.11$^{+0.06,+0.23}_{-0.05,-0.21}$  & 0.51 & 0.16 & 0.399 \\
\enddata
\tablecomments{Here a Sp55 IMF and a Calzetti extinction law is used for
  creating the template. Only ground-based photometric points were used in the
  fits.}
\end{deluxetable}

To check for consistency we used the Bayesian photometric redshift code
\citep{benitez00} to estimate the photometric redshift. This code uses
empirical spectral templates for the fits \citep{coleman80}.  Using the same
ground-based photometric points as above we find
$z_{\mathrm{phot,BPZ}}=0.94\pm0.13$ ($\chi^2/\mathrm{d.o.f.}=1.078$), which is
consistent with the results from HyperZ within 1$\sigma$ uncertainties.
Including the HST data points gives $z_{\mathrm{phot,BPZ}}=0.97\pm0.13$
($\chi^2/\mathrm{d.o.f.}=2.457$). The uncertainties reported here are
1$\sigma$ levels.

\section{Discussion}

Based on multi-colour optical and near-IR photometry we have reported the
first precise photometric redshift for a GRB host galaxy without a
spectroscopic redshift. The host is sufficiently bright that the prospects of
obtaining a spectroscopic redshift, confirming or refuting our proposed value,
appear feasible. We note however that we have been unable to detect emission
lines in public VLT spectra in the ESO archive. In this connection it is
interesting to note the relatively large mean age of the GRB 981226 host
galaxy. This may indicate that emission lines are not prominent in this galaxy
and reminds us that GRB host samples with spectroscopic redshifts may be
biased towards emission line galaxies.  The VLT spectra were obtained at a
relatively low resolution ($>$10 {\AA}), and higher resolution data should be
obtained to test the photometric redshift. At $z\approx 1$ the optical
emission lines could well be contaminated by residuals from subtraction of the
skylines.  Instruments such as ESI, MIKE, or the next generation instrument
X-shooter could reveal such cases.

In addition to the photometric redshift, the SED also allows us to infer
physical properties of the host galaxy.  Assuming a redshift $z=1.11$ and the
currently favored flat cosmology with $\Omega_m=0.3$ and
$H_0=70$~km~s$^{-1}$~Mpc$^{-1}$, the luminosity distance is
$2.3\times10^{28}$~cm.  By interpolating a power law function between the
observed $B$ and $R$ band AB magnitudes (corrected for extinction), we find
that the rest frame 2800~{\AA} Ultra Violet (UV) flux is
0.27$\pm$0.06~$\mu$Jy.  This corresponds to a star-formation rate (SFR) of
1.2$\pm$0.3 M$_{\sun}$~yr$^{-1}$ using the conversion between UV flux and a
global SFR \citep{kennicutt98}. The absolute magnitude of the host galaxy is
$M_B=-20.25$ mag, which implies a specific SFR of 2.4$\pm$0.6
M$_{\sun}$~yr$^{-1}$~$(L/L^*)^{-1}$, where $L^*$ corresponds to the luminosity
of a galaxy with $M_B^*=-21$ mag. This specific star formation rate is smaller
that the UV based SFRs in a sample of 10 GRB hosts \citep{lise04b}.  As for
other GRBs there is no indication that the host galaxy is strongly affected by
extinction.

GRB~981226 had the interesting properties of being consistent with being both
an almost-dark GRB \citep{jakobsson04} and an X-ray flash
\citep{frontera00}\footnote{BATSE detected a short GRB ($T_{90}$=1.7~s, BATSE
  trigger number 7281) which has been associated with GRB 981226, but the
  BATSE burst was triggered one hour after the BeppoSAX detection. Hence,
  these two bursts are different, and the peak energy
  $E_{\mathrm{peak}}=148\pm28$ keV reported in \citet{ghirlanda04c} refers to
  the BATSE burst.}.  We can use the inferred $z_{\mathrm{phot}}$ to estimate
the peak energy for the BeppoSAX burst.  Integrating a Band function and
adding four power law functions to represent the X-ray to $\gamma$-ray
spectrum at different time segments with the parameters given in
\citet{frontera00}, the total time integrated fluence is
$1.9\times10^{-6}$~erg~cm$^{-2}$ and at $z=1.11$ the isotropic energy release
is $E_{\mathrm{iso}}=5.9\times10^{51}$ erg.  \citet{frontera00} divided the
light curve into 5 segments, one of which had
$E_{\mathrm{peak}}^{\mathrm{obs}}$= 61$\pm$15 keV, the others less than
10~keV.  The Amati relation \citep{ghirlanda04} predicts $E_\mathrm{peak}=
77.6$~keV and $E_{\mathrm{peak}}^{\mathrm{obs}} =36.9$ keV, consistent within
2$\sigma$ with the BeppoSAX observations of the peak energy.  With small
uncertainties in the luminosity distances, accurate photometric redshifts for
other GRB hosts will help to constrain cosmological parameters through the
Ghirlanda relation.  The photometric redshift is higher than those reported
for other XRFs but still lower than the median redshifts for GRBs, consistent
with the hypothesis that XRFs may be off-axis GRBs or dirty fireballs which
both predict lower mean redshifts.

\acknowledgments We are very indebted to Dr. E. Harlaftis (\dag\ 2005 Feb 13)
who took some of the VLT optical images analysed in this paper.  We thank Dr.
Micol Bolzonella for assistance with HyperZ. The authors acknowledge benefits
from collaboration within the EU FP5 Research Training Network ``Gamma Ray
Bursts: An Enigma and a Tool''.  L.~Christensen acknowledges support by the
German Verbundforschung associated with the ULTROS project, grant no.
05AE2BAA/4. The Dark Cosmology Centre is supported by the DNRF.  The research
of J.~Gorosabel is partially supported by the Spanish Ministry of Science and
Education through programmes ESP2002-04124-C03-01 and AYA2004-01515 (including
FEDER funds).

%\bibliography{grb981226_ms}

\begin{thebibliography}{42}
\expandafter\ifx\csname natexlab\endcsname\relax\def\natexlab#1{#1}\fi

\bibitem[{{Ben{\'{\i}}tez}(2000)}]{benitez00}
{Ben{\'{\i}}tez}, N. 2000, \apj, 536, 571

\bibitem[{{Bloom} {et~al.}(1998){Bloom}, {Gal}, \& {Meltzer}}]{gcn182}
{Bloom}, J.~S., {Gal}, R.~R., \& {Meltzer}, J. 1998, GRB Circular Network, 182

\bibitem[{{Bloom} {et~al.}(2002){Bloom}, {Kulkarni}, \& {Djorgovski}}]{bloom00}
{Bloom}, J.~S., {Kulkarni}, S.~R., \& {Djorgovski}, S.~G. 2002, AJ, 123, 1111

\bibitem[{{Bolzonella} {et~al.}(2000){Bolzonella}, {Miralles}, \& {Pell{\'
  o}}}]{bolzo00}
{Bolzonella}, M., {Miralles}, J.-M., \& {Pell{\' o}}, R. 2000, A\&A, 363, 476

\bibitem[{{Bruzual} \& {Charlot}(1993)}]{bru93}
{Bruzual}, A.~G., \& {Charlot}, S. 1993, ApJ, 405, 538

\bibitem[{{Bruzual} \& {Charlot}(2003)}]{bruzual03}
---. 2003, \mnras, 344, 1000

\bibitem[{{Calzetti} {et~al.}(2000){Calzetti}, {Armus}, {Bohlin}, {Kinney},
  {Koornneef}, \& {Storchi-Bergmann}}]{calz00}
{Calzetti}, D., {Armus}, L., {Bohlin}, R.~C., {Kinney}, A.~L., {Koornneef}, J.,
  \& {Storchi-Bergmann}, T. 2000, ApJ, 533, 682

\bibitem[{{Castro-Tirado} {et~al.}(1998){Castro-Tirado}, {Gorosabel}, {Drory},
  {Konig}, {Motta}, {Gonzalez}, {Aceituno}, {Greiner}, {Frontera}, \&
  {Palazzi}}]{gcn173}
{Castro-Tirado}, A.~J. {et~al.} 1998, GRB Circular Network, 173

\bibitem[{{Chabrier}(2003)}]{chabrier03}
{Chabrier}, G. 2003, \pasp, 115, 763

\bibitem[{{Christensen} {et~al.}(2004{\natexlab{a}}){Christensen}, {Hjorth}, \&
  {Gorosabel}}]{lise04b}
{Christensen}, L., {Hjorth}, J., \& {Gorosabel}, J. 2004{\natexlab{a}}, \aap,
  425, 913

\bibitem[{{Christensen} {et~al.}(2004{\natexlab{b}}){Christensen}, {Hjorth},
  {Gorosabel}, {Vreeswijk}, {Fruchter}, {Sahu}, \& {Petro}}]{lise04}
{Christensen}, L., {Hjorth}, J., {Gorosabel}, J., {Vreeswijk}, P., {Fruchter},
  A., {Sahu}, K., \& {Petro}, L. 2004{\natexlab{b}}, A\&A, 413, 121

\bibitem[{{Coleman} {et~al.}(1980){Coleman}, {Wu}, \& {Weedman}}]{coleman80}
{Coleman}, G.~D., {Wu}, C.-C., \& {Weedman}, D.~W. 1980, \apjs, 43, 393

\bibitem[{{Fitzpatrick}(1986)}]{fitz86}
{Fitzpatrick}, E.~L. 1986, AJ, 92, 1068

\bibitem[{{Frail} {et~al.}(1999){Frail}, {Kulkarni}, {Bloom}, {Djorgovski},
  {Gorjian}, {Gal}, {Meltzer}, {Sari}, {Chaffee}, {Goodrich}, {Frontera}, \&
  {Costa}}]{frail99}
{Frail}, D.~A. {et~al.} 1999, ApJL, 525, L81

\bibitem[{{Frontera} {et~al.}(2000){Frontera}, {Antonelli}, {Amati},
  {Montanari}, {Costa}, {Dal Fiume}, {Giommi}, {Feroci}, {Gennaro}, {Heise},
  {Masetti}, {Muller}, {Nicastro}, {Orlandini}, {Palazzi}, {Pian}, {Piro},
  {Soffitta}, {Stornelli}, {in 't Zand}, {Frail}, {Kulkarni}, \&
  {Vietri}}]{frontera00}
{Frontera}, F. {et~al.} 2000, \apj, 540, 697

\bibitem[{{Galama} {et~al.}(1998){Galama}, {Vreeswijk}, {Palazzi}, {Frontera},
  {Masetti}, {van Paradijs}, {Kouveliotou}, \& {Collier Cameron}}]{gcn172}
{Galama}, T.~J., {et~al.} 1998, GRB Circular Network, 172

\bibitem[{{Gehrels} {et~al.}(2004){Gehrels}, {Chincarini}, {Giommi}, {Mason},
  {Nousek}, {Wells}, {White}, {Barthelmy}, {Burrows}, {Cominsky}, {Hurley},
  {Marshall}, {M{\' e}sz{\' a}ros}, {Roming}, {Angelini}, {Barbier}, {Belloni},
  {Campana}, {Caraveo}, {Chester}, {Citterio}, {Cline}, {Cropper}, {Cummings},
  {Dean}, {Feigelson}, {Fenimore}, {Frail}, {Fruchter}, {Garmire}, {Gendreau},
  {Ghisellini}, {Greiner}, {Hill}, {Hunsberger}, {Krimm}, {Kulkarni}, {Kumar},
  {Lebrun}, {Lloyd-Ronning}, {Markwardt}, {Mattson}, {Mushotzky}, {Norris},
  {Osborne}, {Paczynski}, {Palmer}, {Park}, {Parsons}, {Paul}, {Rees},
  {Reynolds}, {Rhoads}, {Sasseen}, {Schaefer}, {Short}, {Smale}, {Smith},
  {Stella}, {Tagliaferri}, {Takahashi}, {Tashiro}, {Townsley}, {Tueller},
  {Turner}, {Vietri}, {Voges}, {Ward}, {Willingale}, {Zerbi}, \&
  {Zhang}}]{gehrels04}
{Gehrels}, N. {et~al.} 2004, \apj, 611, 1005

\bibitem[{{Ghirlanda} {et~al.}(2004{\natexlab{a}}){Ghirlanda}, {Ghisellini}, \&
  {Lazzati}}]{ghirlanda04}
{Ghirlanda}, G., {Ghisellini}, G., \& {Lazzati}, D. 2004{\natexlab{a}}, \apj,
  616, 331

\bibitem[{{Ghirlanda} {et~al.}(2004{\natexlab{b}}){Ghirlanda}, {Ghisellini}, \&
  {Celotti}}]{ghirlanda04c}
{Ghirlanda}, G., {Ghisellini}, G., \& {Celotti}, A. 2004{\natexlab{b}}, \aap,
  422, L55

\bibitem[{{Gorosabel} {et~al.}(2003{\natexlab{a}}){Gorosabel}, {Christensen},
  {Hjorth}, {Fynbo}, {Pedersen}, {Jensen}, {Andersen}, {Lund}, {Jaunsen},
  {Castro Cer{\' o}n}, {Castro-Tirado}, {Fruchter}, {Greiner}, {Pian},
  {Vreeswijk}, {Burud}, {Frontera}, {Kaper}, {Klose}, {Kouveliotou}, {Masetti},
  {Palazzi}, {Rhoads}, {Rol}, {Salamanca}, {Tanvir}, {Wijers}, \& {van den
  Heuvel}}]{goro03b}
{Gorosabel}, J. {et~al.} 2003{\natexlab{a}}, \aap, 400, 127

\bibitem[{{Gorosabel} {et~al.}(2003{\natexlab{b}}){Gorosabel}, {Klose},
  {Christensen}, {Fynbo}, {Hjorth}, {Greiner}, {Tanvir}, {Jensen}, {Pedersen},
  {Holland}, {Lund}, {Jaunsen}, {Castro Cer{\' o}n}, {Castro-Tirado},
  {Fruchter}, {Pian}, {Vreeswijk}, {Burud}, {Frontera}, {Kaper}, {Kouveliotou},
  {Masetti}, {Palazzi}, {Rhoads}, {Rol}, {Salamanca}, {Wijers}, \& {van den
  Heuvel}}]{goro03}
---. 2003{\natexlab{b}}, \aap, 409, 123

\bibitem[{{Gorosabel} {et~al.}(2005){Gorosabel}, {Perez-Ramirez}, {Sollerman},
  {de Ugarte-Postigo}, {Fynbo}, {Castro-Tirado}, {Jakobsson}, {Christensen},
  {Hjorth}, {Johanneson}, {Guizy}, {Castro-Ceron}, {Bjornsson}, {Sokolov},
  {Fatkhullin} \& {Nilson}}]{goro05}
{Gorosabel}, J. {et~al.} 2005, A\&A in press (astro-ph/0507488)

\bibitem[{{Holland}(2000)}]{holland_gcn}
{Holland}, S. 2000, GRB Circular Network, 749

\bibitem[{{Holland}(2001)}]{holland01q}
{Holland}, S. 2001, in AIP Conf. Proc. 586: 20th Texas Symposium on
  relativistic astrophysics, 593

\bibitem[{{Jakobsson} {et~al.}(2004){Jakobsson}, {Hjorth}, {Fynbo}, {Watson},
  {Pedersen}, {Bj{\" o}rnsson}, \& {Gorosabel}}]{jakobsson04}
{Jakobsson}, P., {Hjorth}, J., {Fynbo}, J.~P.~U., {Watson}, D., {Pedersen}, K.,
  {Bj{\" o}rnsson}, G., \& {Gorosabel}, J. 2004, \apjl, 617, L21

\bibitem[{{Jakobsson} {et~al.}(2005){Jakobsson}, {Bj{\"
  o}rnsson}, {Fynbo}, {J\'ohanneson}, {Hjorth}, {Thomsen}, {M{\o}ller},
  {Watson}, \& {Jensen}}]{jakobsson05b}
{Jakobsson}, P. {et~al.} 2005, MNRAS in press (astro-ph/0505542)

\bibitem[{{Kennicutt}(1998)}]{kennicutt98}
{Kennicutt}, R.~C. 1998, ARA\&A, 36, 189

\bibitem[{{Kinney} {et~al.}(1996){Kinney}, {Calzetti}, {Bohlin}, {McQuade},
  {Storchi-Bergmann}, \& {Schmitt}}]{kinney96}
{Kinney}, A.~L., {Calzetti}, D., {Bohlin}, R.~C., {McQuade}, K.,
  {Storchi-Bergmann}, T., \& {Schmitt}, H.~R. 1996, ApJ, 467, 38

\bibitem[{{Klose}(1998)}]{gcn186}
{Klose}, S. 1998, GRB Circular Network, 186

\bibitem[{{Le Floc'h} {et~al.}(2003){Le Floc'h}, {Duc}, {Mirabel}, {Sanders},
  {Bosch}, {Diaz}, {Donzelli}, {Rodrigues}, {Courvoisier}, {Greiner},
  {Mereghetti}, {Melnick}, {Maza}, \& {Minniti}}]{lefloch03}
{Le Floc'h}, E. {et~al.} 2003, \aap, 400, 499

\bibitem[{{Lindgren} {et~al.}(1999){Lindgren}, {Hjorth}, {Pedersen},
  {Andersen}, {Jaunsen}, {Sollerman}, {Smoker}, {Mooney}, \&
  {Palazzi}}]{gcn190}
{Lindgren}, B. {et~al.} 1999, GRB Circular Network, 190

\bibitem[{{Miller} \& {Scalo}(1979)}]{miller79}
{Miller}, G.~E., \& {Scalo}, J.~M. 1979, ApJS, 41, 513

\bibitem[{{Prevot} {et~al.}(1984){Prevot}, {Lequeux}, {Prevot}, {Maurice}, \&
  {Rocca-Volmerange}}]{prevot84}
{Prevot}, M.~L., {Lequeux}, J., {Prevot}, L., {Maurice}, E., \&
  {Rocca-Volmerange}, B. 1984, A\&A, 132, 389

\bibitem[{{Rhoads} {et~al.}(1998){Rhoads}, {Orosz}, {Lee}, \&
  {Stassun}}]{gcn181}
{Rhoads}, J., {Orosz}, J.~A., {Lee}, J., \& {Stassun}, K. 1998, GRB Circular
  Network, 181

\bibitem[{{Rol} {et~al.}(2005){Rol}, {Wijers}, {Kouveliotou}, {Kaper}, \&
  {Kaneko}}]{rol05}
{Rol}, E., {Wijers}, R.~A.~M.~J., {Kouveliotou}, C., {Kaper}, L., \& {Kaneko},
  Y. 2005, \apj, 624, 868

\bibitem[{{Salpeter}(1955)}]{salpeter55}
{Salpeter}, E.~E. 1955, \apj, 121, 161

\bibitem[{{Schaefer} {et~al.}(1998){Schaefer}, {Kemp}, {Feygina}, \&
  {Halpern}}]{gcn185}
{Schaefer}, B.~E., {Kemp}, J., {Feygina}, I., \& {Halpern}, J. 1998, GRB
  Circular Network, 185

\bibitem[{{Schlegel} {et~al.}(1998){Schlegel}, {Finkbeiner}, \&
  {Davis}}]{schlegel98}
{Schlegel}, D.~J., {Finkbeiner}, D.~P., \& {Davis}, M. 1998, ApJ, 500, 525

\bibitem[{{Seaton}(1979)}]{seaton79}
{Seaton}, M.~J. 1979, MNRAS, 187, 73P

\bibitem[{{Taylor} {et~al.}(1998){Taylor}, {Frail}, {Kulkarni}, {Shepherd},
  {Feroci}, \& {Frontera}}]{taylor98}
{Taylor}, G.~B., {Frail}, D.~A., {Kulkarni}, S.~R., {Shepherd}, D.~S.,
  {Feroci}, M., \& {Frontera}, F. 1998, \apjl, 502, L115

\bibitem[{{Wozniak}(1998)}]{gcn177}
{Wozniak}, P.~R. 1998, GRB Circular Network, 177

\end{thebibliography}

\end{document}